\documentstyle[natbib,aas2pp4]{article}
\slugcomment{To appear in  Matter in the
Universe, Space Sciences Series of ISSI, Kluwer
Academic, Dordrecht, and in Space Science Reviews, 
eds. P. Jetzer, K. Pretzl and R. von Steiger, } \lefthead{J.
Silk} \righthead{Supermassive Black Holes and Galaxy Formation}

\date{\today}
\begin{document}

\title{{\bf Supermassive Black Holes and Galaxy Formation}
\footnote{To appear in  Matter in the
Universe, Space Sciences Series of ISSI, Kluwer
Academic, Dordrecht, and in Space Science Reviews, 
eds. P. Jetzer, K. Pretzl and R. von Steiger, } 
}
\author{ JOSEPH SILK
%\altaffilmark{1}
}
%\and
%\author{JOSEPH  SILK}
\affil{Department of Physics, Nuclear and Astrophysics
                Laboratory, Keble Road, Oxford OX1 3RH, United Kingdom}
%\altaffiltext{1}{Inquiries can be sent to {\bf
%lopes@astro.ox.ac.uk}}
%\altaffiltext{2}{Instituto Superior
%T\'ecnico, Av. Rovisco Pais,
%                Centro Multidisciplinar de Astrof\'\i sica,
%                1049-001 Lisboa, Portugal}

\setcounter{page}{1}

%% Put any additional command definitions here
\newcommand{\RON}{read-out noise}
\newcommand{\beqa}{\begin{eqnarray}}
\newcommand{\eeqa}{\end{eqnarray}}
%\begin{document}
\newcommand{\tnew}[1]{{\bf #1}}

%% Do not remove the following six lines:

%% Title - should be in capitals:

%\maketitle
%#################################################################
%
\begin{abstract}
The formation of supermassive black holes (SMBH)
is intimately related to galaxy formation, although 
precisely how remains a mystery.
I speculate that formation of, and feedback from, SMBH may alleviate problems that have arisen in our understanding of the cores of dark halos of galaxies.

\end{abstract}
%
%\keywords{Key~words: stars: oscillations - stars: interiors - Sun:
%oscillations - Sun: interior: cosmology - dark matter}

\section{INTRODUCTION}

Galaxy formation theory is not in a  very satisfactory state. This stems
ultimately from our lack of any fundamental understanding of star
formation. There is no robust theory for the detailed properties of
galaxies.
In contrast, the hierarchical 
formation of large-scale structure in a cold dark matter-dominated
Friedmann-Lema\^itre universe has successfully confronted essentially all
observations, ranging from deep surveys of the galaxy distribution, the
formation of galaxy clusters, and the temperature fluctuations in the cosmic
microwave background.

\section{The apparent demise of CDM}

The advent of high resolution N-body simulations has revealed challenges for
galaxy formation.  Dark halos contain considerable sub-structure, amounting 
to of order 10\%  of the halo mass, and continuing down to unresolved scales
of $10^6\rm M_\odot$ or smaller. 
This has led to two problems. One is that the predicted number of dwarf
gallaxy satellites exceeds that observed around the Milky Way
by an order of magnitude.  Even very low surface brightness  dwarfs are
easily
observed in our local environment, and efficient mass loss prior to star
formation has been  invoked  to resolve this discrepancy (Gnedin 2001). 

However, since stars
will form before the gas is entirely stripped,
the dwarfs should be visible. Star formation is a local process that
occurs 
in localized inhomogeneities, and
the local
free-fall time is short because the sound crossing time is short, whereas
stripping is a global process that requires star formation to first occur,
even if externally imposed,  e.g.  by an ionizing UV radiation field.

Another manifestation  of excessive  substructure in CDM halos is the failure 
of the CDM model to account for  the velocity dispersion distribution  
of low ionization damped Lyman alpha systems at $z>1.5$ (Prochaska and Wolfe 2001).
The kinematics reveal too few clouds with velocity widths of 100 km/s or more
compared to the number of low velocity width systems.
It is unlikely that feedback plays much of a role in these systems, selected
by absorption in low ionization, neutral gas, and with velocity widths that
are unlikely to be affected by photoionization.

A second problem is that while the initial angular momentum of a typical
protogalaxy  when the halo first collapsed   can reproduce the observed size
distribution of disks if angular momentum
is approximately conserved during  baryon infall and disk accretion, the
numerical
simulations show that most of the angular momentum is actually lost to the
dark halo.
The clumpiness induces strong angular momentum transfer via tidal torquing
and dynamical friction from the dissipating baryons to the energy-conserving
dark matter. The resulting disks have too little specific 
angular momentum   by an order of magnitude 
%.g.\cite{nava,eke}
(e.g., Navarro and Steinmetz 2000a; Eke, Navarro and Steinmetz 2001).
It seems likely that  stellar feedback  can in principle provide sufficient 
input of momentum to the infalling baryons via supernova explosions and
remnant formation to avoid much of the angular momentum loss, at the price
however of a significant delay in disk formation. Giant disks, in particular,
would form relatively late, in possible conflict both
with the observational
evidence at $z\sim 1$ (Faber et al. 2001) and with ages in the outer parts of nearby disks 
(Ferguson and Johnson 2001).
A more serious concern for those theoretically inclined 
is that there is as yet no
robust prescription for the required feedback, although progress is being made (Couchman and Thacker 2001).

Another result from the high resolution simulations of galaxy halos is that 
the dark halo has a central concentration and a central cusp.
The halo profile is described by a universal profile
$$\rho = \frac{A}{r^\gamma(r+r_s)^{3-\gamma}},$$
where $A$ and $r_s$ depend on the cosmological model parameters and
normalization, and $1<\gamma< 1.5$.
The CDM concentration parameter, defined as $C=r_s/r_v$ where $r_s$ is the
scale radius where the density profile flattens and defines a cusp or core,
and $r_v$ is the virial radius, is predicted to be 10-20 by numerical
simulations. This implies for example that about half the mass within the
half-light radius (for our galaxy this corresponds to near the solar circle)
is in the form of CDM
(e.g.  Navarro and Steinmetz 2000b; Eke, Navarro and Steinmetz 2001). This
apparently  contradicts bulge microlensing studies for the
Milky Way, which permit a CDM
fraction of at most 10\% within the solar circle, when combined with the
rotation curve and infrared (DIRBE) stellar population modelling (Binney and
Evans 2001).
The  status of the dark matter content of the inner Milky Way is uncertain
by about a factor of 2 due to the uncertainty in the optical depth to microlensing.
Dynamical studies of nearby barred galaxies
conclude that self-gravity of the bars, which consist primarily of stars,
must dominate the inner gravitational potential (Sellwood and Kosowsky 2000).
At most, a 10\% contribution by spherically or axially-symmetrically
distributed dark matter can be allowed within the bar region
to avoid conflict
with the observed kinematics of certain barred
galaxies.
Both the bars and the matter within the inner galaxy out to a bar radius
are required  to be baryon-dominated
by the need for  gravitational instability to
explain the existence of bars. 

In contrast, a recent study of a non-barred spiral finds that detailed modelling of
the rotation curve with a maximal disk requires an axially symmetric
 dark matter contribution of about 30
\%
within an optical radius (Kranz, Slyz and Rix 2001). 
This is not inconsistent with bar formation by secular
instability of a cold disk, but raises  the question of 
why there is  an apparently large
spread in dark matter content among  disk galaxies.

The pioneering study by Navarro, Frenk and White (1997)
found a cusp profile 
of $\gamma \approx 1,$
but this has been superceded by higher resolution simulations which almost
invariably find that $\gamma \approx 1.5$ for galaxy mass halos (Moore et
al. 1999; Jing and Suto 2000;
Klypin et al. 2001).

There is another observational hurdle to overcome. 
Low surface brightness  spirals are everywhere  dark matter-dominated,
and so provide outstanding laboratories for dark matter studies via rotation
curves.
High resolution H$\alpha$  rotation curves reveal  a wide array of central
profiles. Most systems have soft cores without any indication of a central
cusp, but some do show indications of $\gamma$  being as large as
unity. None however approach the predicted value of 1.5, and the best studied
examples are in clear conflict with such a steep cusp (van den Bosch and
Swaters 2001).
 
\section{Resurrection of CDM}

 Cold dark matter is seriously challenged.
Whether it is actually dead is quite another matter. 
Nevertheless, to confront this possibility, there have been numerous attempts
to resurrect CDM. These come under two distinct guises: tinkering with the
particle physics or elaborating on the astrophysics.

Particle physics variations include the introduction of self-interacting dark
matter. Self-interactions, via elastic scattering, allow the dark matter to
develop a smooth core.
If the scattering mean free path is adjusted to be of the order of the core
size, one can avoid
developing an excessive central concentration of dark matter. This comes at a
price, however: for example, the dark halo is found  to be spherical, 
in apparent contrast to the
elongated shapes inferred via gravitational lensing of  dominant
massive cluster galaxy halos (Miralda-Escude 2000).

Warm dark matter has provided a
possible resolution of the cold dark matter
``crisis''. A sterile neutrino with a mass of around 1 keV can constitute the
dark matter, since its sterility enables the neutrino abundance to be
suppressed relative  
to the ordinary neutrinos, and thereby avoid the hot dark matter mass bound
for overclosing the universe. At 1 keV, the free-streaming length is of
order the comoving scale of a dwarf galaxy. Substructure is suppressed.
However high resolution simulations find that warm dark matter still produces
a central cusp, although with $\gamma \approx 1.$
(Bode et al. 2001; Knebe et al. 2001). Possibly more serious
 is another consequence of the suppression of small-scale structure, which
results in late formation of dwarf galaxies. If nearby dwarfs contain
genuinely old stellar populations, this would be difficult to understand in
the context of warm dark matter. Perhaps more serious maybe the need for
a substantial spatial density  of ultraluminous quasars at $z> 6,$ each requiring a
central SMBH of mass somewhat in excess of $10^9\rm M_\odot$ (Fan et al. 2001).
The diminished power of WDM makes 
it difficult to  reconcile  observations of
this type with WDM. The abundance of massive ($\sim 10^{12}\rm M_\odot$)
halos 
is unaffected, but these form late: it is necessary to have small-scale
power at high redshift to form the dense cores within which the SMBH formed. 

More complex versions of particle dark matter have been proposed. These
include self-interacting warm dark matter, a model which inspires little
confidence given the difficulties encountered by warm and interacting dark
matter, and more exotic variants such as shadow dark matter.
Modified particle dark matter can no doubt be developed to explain all of
the required dark matter properties. However the seductive simplicity of the
SUSY LSP as an attractive candidate for CDM is lost.

An alternative approach is via the astrophysics of galaxy formation. Can the
dark matter
profile be modified by astrophysical processes? The answer is perhaps.
Consider the following sequence of events.  Supermassive black holes form at
the centres of dark halos, possibly contemporaneously with, and certainly
coupled to,
the formation of the stellar spheroid. This sequence of events is strongly
motivated by the observed correlation between supermassive black hole mass
and spheroid mass (Ferrarese et al. 2000; Gebhardt et al. 2000)
as well as by the super-solar abundances found in quasar broad emission line
regions (Hamann et al. 2001).

 Now consider a merger between  a dwarf spheroid,
containing a massive black hole, and a much larger galaxy. The smaller 
massive black hole spirals in
under dynamical friction  and eventually merges 
with the SMBH of the dominant galaxy.
The details of the merger are not clear, but it seems likely that the smaller
SMBH  decays into  tighter and
tighter orbits as stars are ejected into the regime where gravitational
radiation eventually takes over.  
Gas is essential for the initial formation of the SMBH, but dark matter cores
and stars are
crucial for the SMBH to undergo merger-induced growth.
The process is well matched to galaxy formation.
The dwarf galaxy stars are stripped and help feed and regulate spheroid
growth.

However there is feedback on the dark matter.
The SMBH merger results in heating of the dark matter cusp.
The region that undergoes heating can be quite extensive as a transient
rapidly rotating gaseous bar forms during the merger, and is slowed
by dynamical friction. The dark matter is heated and acquires angular
momentum.
The result is that the concentration and cusp are likely to be modified. 
Simulations suggest that the inner profile flattens to $\gamma \approx 0.5$
(Nakano and Makino 1999; Merritt, Cruz and Milosavljevic 2001)
This is considerably flatter than the initial cusp, and is insensitive to the
initial density profile. The substructure is unafffected. However the 
angular momentum acquired by the halo will
help in reinjecting angular momentum into infalling gas clouds that 
form the disk over a time-scale of a gigayear or longer.

A more dramatic
interaction of the black hole with the dark matter may be imagined. Suppose
there is an early accretion phase onto the SMBH, perhaps driven by transient
bar formation. The activated SMBH will produce a vigorous outflow. The inner
region is baryon-dominated. If enough baryonic mass loss occurs, the inner
dark matter profile will be less concentrated and more uniform
(Binney, Gerhard and Silk 2001).
 The modifications occur within the region where the baryon content changes
from being dominant to being sub-dominant, i.e. of order 
half of the baryonic mass must be driven out, possibly just into the halo.

LSB dwarfs may be extreme examples where such mass loss has occurred.
It is precisely these objects that provide 
possible evidence for 'discrepant' CDM profiles.
Even in the absence of strong outflows, important dynamical heating can occur
via dynamical friction on the rotating bar. Hence LSB galaxies, galaxies
which had primeval bars
(and so may be vulnerable to formation of a second bar if there is a suitable
supply of gas), and galaxies with soft stellar cores, where SMBH mergers may
have initiated stellar ejection,
 are the prime examples where the initial CDM profile is likely
to have been modified.

Evidently, there is no firm prediction about dark matter profiles, whether 
in LSB galaxies or in luminous galaxies. I turn now to observations of dark
matter, and  discuss whether one can indeed observe the inner profile of
dark halos.

\section{"Observing" dark matter}
The favoured candidate  for  CDM is the lightest stable SUSY relic
particle.
This  must be neutral (to avoid already having been detected)
and its mass  is constrained by accelerator searches and theoretical 
considerations of thermal freeze-out to
lie in the range
50 GeV to a few TeV. The relic density  
is determined when annihilations and pair production go out of thermal
equilibrium  in the early universe
at $T\sim m_x/20k,$ and one infers that $\Omega_x\propto \sigma_{ann}^{-1},$
where $\sigma_{ann}$ is the annihilation cross-section  extrapolated to
the low temperature limit. For typical weak interaction values of
$\sigma_{ann}$, one finds that $\Omega_x\sim 0.3$ is required to  account
for the dark matter content of the universe. Via 
studying a grid  of SUSY models, one can infer 
a range of particle masses from the annihilation cross-section. Were it
not for the accelerator bounds on the sparticle masses, the uncertainty 
in $m_x$ would span some 5 orders of magnitude.

The annihilation cross-section and  particle mass is constrained.
So also is the elastic scattering cross-section once the annihilation
cross-section is specified.
 This means that  one can now consider  possible detection schemes.
The obvious one is direct detection by elastic scattering.
Use of annual modulation of the incident flux on a terrestrial detector 
has led to a tentative detection  (DAMA) that requires an implausibly
large cross-section  given the suite of minimal SUSY models
and is marginally inconsistent with another experiment  (CDMS).
Annihilations result in hadronic jets that decay into gamma rays, high
energy electron-positron pairs, proton-antiproton pairs and neutrinos, all
of which are potentially detectable as galactic halo signals.
The EGRET gamma ray detector on CGRO has reported 
a diffuse high latitude gamma ray flux
that can only be accounted for by annihilations if the halo is clumpy by a
factor $<n^2>/<n>^2\sim 100$ (Calcaneo-Roldan and Moore 2000).
In fact, the observed signal has a spectral  signature that does not
resemble that expected for annihilations. Presumably it is due to
unresolved distant sources.

Another possible signature of CDM is associated with  a feature
in the cosmic ray positron spectrum at very high energy.
The HEAT experiment reported a feature near 100 GeV that, if real,
cannot be explained by secondary interactions between cosmic rays and the
interstellar medium. Annihilation products of a  100 GEV neutralino would
provide an excellent fit
to the data, except that one requires the annihilation cross-section to be
boosted by a factor $\sim 100$ relative to typical models.
Clumpiness in the dark halo again provides a possible explanation in terms
of a WIMP signal (Baltz and Edsjo 1999).

Perhaps the most exciting prospect comes from the Galactic Centre, where
annihilations may already have been seen. The supermassive black hole at
the Galactic Centre of $2.6\times10^6\rm M_\odot$ most likely
formed by baryonic
dissipation within the already existing dark halo.
If the growth process is approximately adiabatic, the neutralinos form a
central cusp with slope $\rho\propto r^{-\gamma'},$ where
$\gamma'=\frac{9-2\gamma}{4-\gamma},$ and the central dark
halo cusp slope
is $\gamma.$ High resolution halo simulations suggest that $\gamma\approx
1.5,$ but  black hole merging softens the initial CDM cusp
to $\gamma\approx 0.5.$ Even if the cusp were initially destroyed and were 
isothermal, $  \gamma\approx 0$, the spike has $\gamma'> 1.5$ and the
annihilations therefore diverge within
the zone of influence of the black hole, at a radius $\sim
GM_{bh}/\sigma^2\sim \rm 0.1 \; pc,$ down to about 10 Schwarzschild radii,
$\sim 10^{-6}\rm pc.$

Signals from the enhanced annihilations may already have been detected
(Bertone, Sigl and Silk 2001).
The radio flux from Sag A*, the  
unidentified source at the Galactic  Centre, can be accounted for in
spectral shape by the annihilation signal from electron-positron pairs
undergoing synchrotron radiation, which is self-absorbed.
The normalization depends on what assumes about the magnetic field near the
SMBH as well as on the central cusp profile.
One can eliminate the uncertainty in modelling the magnetic field by 
calculating the flux of 
gamma rays, and 
the spectral distribution of the
EGRET gamma ray flux from the unresolved source at the Galactic Centre 
can be explained. To simultaneously account for the gamma ray flux as well as
the synchrotron flux from Sag A*, it is necessary to adopt 
 a magnetic field that is below the equipartition value by a factor of
10 or so.

All of this is necessarily highly speculative. Future observations may
 greatly help in 
pinning down the CDM  characterstics. Annihilations also generate high
 energy neutrinos.
These  propagate freely from the vicinity of the SMBH at the Galactic
 Centre, and detection  would provide unambiguous support for annihilating
 WIMPs.
The predicted fluxes are within the anticipated sensitivity of the ANTARES
neutrino detector, now under construction.

In summary, supermassive black holes, for better or for worse, are intimately
 connected with the
process of galaxy spheroid formation. Whether they aid and abet formation of
 the first stars remains a mystery. There are certainly dynamical
and most likely astrochemical links. Given the unabated array  of
challenges that CDM is facing in its canonical version, as formulated by
 so-called semi-analytical galaxy formation,
it is tempting to appeal to a totally new ingredient in formulating galaxy
 formation theory to help resolve these issues. It remains to be seen whether
 the ultimate answer lies in the dynamical feedback of SMBH formation and
 evolution on dark halo cores, or on a new prescription that modifies the
 physics of CDM, or possibly in fundamental physics whereby  on
large scales   unanticipated changes in  4-d Einstein gravity may be appearing,
such as might be associated with the influence of higher dimensions. 
My preference is for the first of these alternatives, but observations will
 be the ultimate arbiter.

 %\section*{ACKNOWLEDGMENTS}

%\section*{References}

\bibliographystyle{dmprl}
%\bibliography{helioseismology}

%\clearpage \onecolumn
%%%%%%%%%%%%%%%%%%%%%%%%%%%%%%%%%%%%%%%%%%%%%%%%%%%%%%%%%%%%%%%%%%%%%%%%%%%%%

\end{document}